\documentclass[12pt,preprint]{aastex}

\begin{document}

\title{The Low Frequency Radio Counterpart of the XMM \\
Large Scale Structure Survey}

\author{A.~S.~Cohen \altaffilmark{1,3},
H.~J.~A.~R\"ottgering \altaffilmark{2}, 
N.~E.~Kassim \altaffilmark{1},
W.~D.~Cotton \altaffilmark{5},
R.~A.~Perley \altaffilmark{4},
R.~Wilman \altaffilmark{2}, 
P.~Best \altaffilmark{7},
M.~Pierre \altaffilmark{6},
M.~Birkinshaw \altaffilmark{8},
M.~Bremer \altaffilmark{8},
and A.~Zanichelli \altaffilmark{9}
}

\altaffiltext{1}{Naval Research Laboratory, Code 7213, Washington, DC, 20375 USA, cohen@rsd.nrl.navy.mil, kassim@rsd.nrl.navy.mil} 
\altaffiltext{2}{Leiden Univertity, Sterrewacht, Oort Gebouw, P.O. Box 9513, 
2300 RA Leiden, The Netherlands, rottgeri@strw.LeidenUniv.nl}
\altaffiltext{3}{National Research Council Postdoctoral Fellow}
\altaffiltext{4}{National Radio Astronomy Observatory, P.O. Box 0, Socorro, NM 87801 USA}
\altaffiltext{5}{National Radio Astronomy Observatory, 520 Edgemont Road, Charlottesville, VA, 22903 USA}
\altaffiltext{6}{Saclay}
\altaffiltext{7}{Institute for Astronomy, Royal Observatory Edinburgh, 
Blackford Hill, Edinburgh, EH9 3HJ, UK}
\altaffiltext{8}{Department of Physics, University of Bristol, Tyndall Avenue, Bristol BS8 1TL, UK}
\altaffiltext{9}{Istituto di Radioastronomia - CNR, Bologna, Italy}

\begin{abstract}

The XMM Large Scale Structure Survey (XMM-LSS) is a major project
to map the large scale structure of the universe out to 
cosmological distances.  An $8^{\circ}\times8^{\circ}$ region will 
be surveyed by XMM with planned optical follow-up to produce a 
three-dimensional map of many hundreds of clusters out to a redshift 
of $z = 1$.  To explore the relation of the large scale structure to
the location and properties of extragalactic radio sources, the XMM-LSS
project also includes a low frequency radio survey of this region.  
This combination will provide unprecedented insight into how the radio 
source formation and evolution are affected by the local environment.   
Here, we present preliminary results from our 325 MHz and 
74 MHz surveys in this region.  
At 325 MHz, we have a flux limit of 4 mJy/beam, a resolution of 
$6.3''$, and a total of 256 source detections over 5.6 square 
degrees.
At 74 MHz, we have a flux limit of 275 mJy/beam, a resolution of 
$30''$, and a total of 211 source detections over 110 square degrees.
We describe these results and explore what they tell us about the
population of extra-galactic low frequency radio sources.  
The 74 MHz survey represents the first 
presentation of a deep, sub-arcminute resolution survey at such a low 
frequency.  
This was made possible by recent 
advances in both hardware and data reduction algorithms which we 
describe in detail.

\end{abstract}

\section{Introduction}

The XMM Large Scale Structure Survey (XMM-LSS) is a deep X-ray survey
designed to map the distribution of groups and clusters of galaxies
out to a redshift of $z = 1$ \citep{2001Msngr.105...32P}.  This will 
provide an unprecedented 
view of the large scale structure of the universe.  For the first time, 
the cluster correlation function will be measured for high redshifts.  
The planned optical follow-up will compare the spatial distribution of 
AGNs with this large scale structure, and study the origin
of AGN/QSO and how it relates to its surroundings.  Comparison of the
cosmic web determined from X-rays with the galaxy distribution determined
optically will shed light on bias mechanisms as a function of redshift.

The XMM-LSS survey consists of a $24\times24$ grid of XMM pointings, each 
with 10 ks integration, that will uniformly map a square 
$8^{\circ}\times8^{\circ}$ region to a sensitivity for point-like sources
of about $5\times10^{-15}\mbox{erg}\,\mbox{cm}^{-2}\,\mbox{s}^{-1}$ in 
the [0.5-2] keV band.  
A thorough study of the expected number of clusters detected by this 
survey was recently carried out by \citet{2002A&A...390....1R}.  Using
the non-evolving luminosity temperature (L-T) relation of 
\citet{1999MNRAS.305..631A}, the Press-Schechter formalism 
\citep{1974ApJ...187..425P} and X-ray image simulations, the number
of detectable clusters was predicted as a function of redshift 
for various cosmologies.  In the favored $\Lambda$CDM cosmology, they 
predict that XMM-LSS should detect about 600-1200 clusters within 
$0 < z < 1$ and another approximately 300 at $1 < z < 2$. 

An optical follow-up in the XMM-LSS region is planned with the MEGACAM 
at the CFHT.  Deep ($I \approx 25$) imaging of the entire region will 
detect a sufficient number density of galaxies to allow for weak lensing 
detections of large filaments.  A planned spectroscopic follow-up 
program will measure the redshifts for all candidate clusters identified 
in the X-ray survey.

With the prospect of these data, we conducted a low-frequency VLA survey 
of the XMM-LSS region in order to investigate the relationship between 
the population of cosmological radio sources and the large scale structure 
of the universe.  Specific questions we wish to address include: (i) 
Where are various populations of radio sources and X-ray quasars located 
with respect to the LSS? (ii) How does the environment of radio sources 
influence their fundamental physical properties such as linear size and 
radio power? (iii) What is the connection between distant Mpc-sized radio 
halos and the dynamical state of their associated clusters?

Our low-frequency radio survey was conducted with the VLA in the A-array 
configuration and mapped the entire XMM-LSS region at 74 MHz and 5.6 
square degrees of this region at 325 MHz.  This was made possible by 
recent advances in low-frequency imaging including the addition of 74 MHz 
receivers to the VLA in 1997, and the development of new software and data 
reduction algorithms which make it possible to map the entire primary beam 
at these frequencies.  

At 325 MHz our limiting flux density ($5\sigma_{rms}$ level, where 
$\sigma_{rms} \equiv$ map rms) is 4.0 mJy and our resolution is $6.3''$.  
For comparison, the WENSS survey \citep{1997A&AS..124..259R},
conducted at 325 MHz and 352 MHz, which covers the northern sky above 
$\delta = +30^{\circ}$ has a limiting 
flux density of about 18 mJy/beam with a resolution of 
about an arcminute.  A deeper survey of a smaller area (95 square degrees)
at 327 MHz has been done at Westerbork \citep{1993BICDS..43...17W}, 
reaching about 3 mJy/beam over 95 square degrees, though also at one 
arcminute resolution.  The 365 MHz Texas survey \citep{1996AJ....111.1945D}
covers the sky between $-35.5^{\circ} \leq \delta \leq +71.5^{\circ}$, but 
only reaches a depth of 250 mJy/beam.  

\begin{figure}
\epsscale{0.75}
\plotone{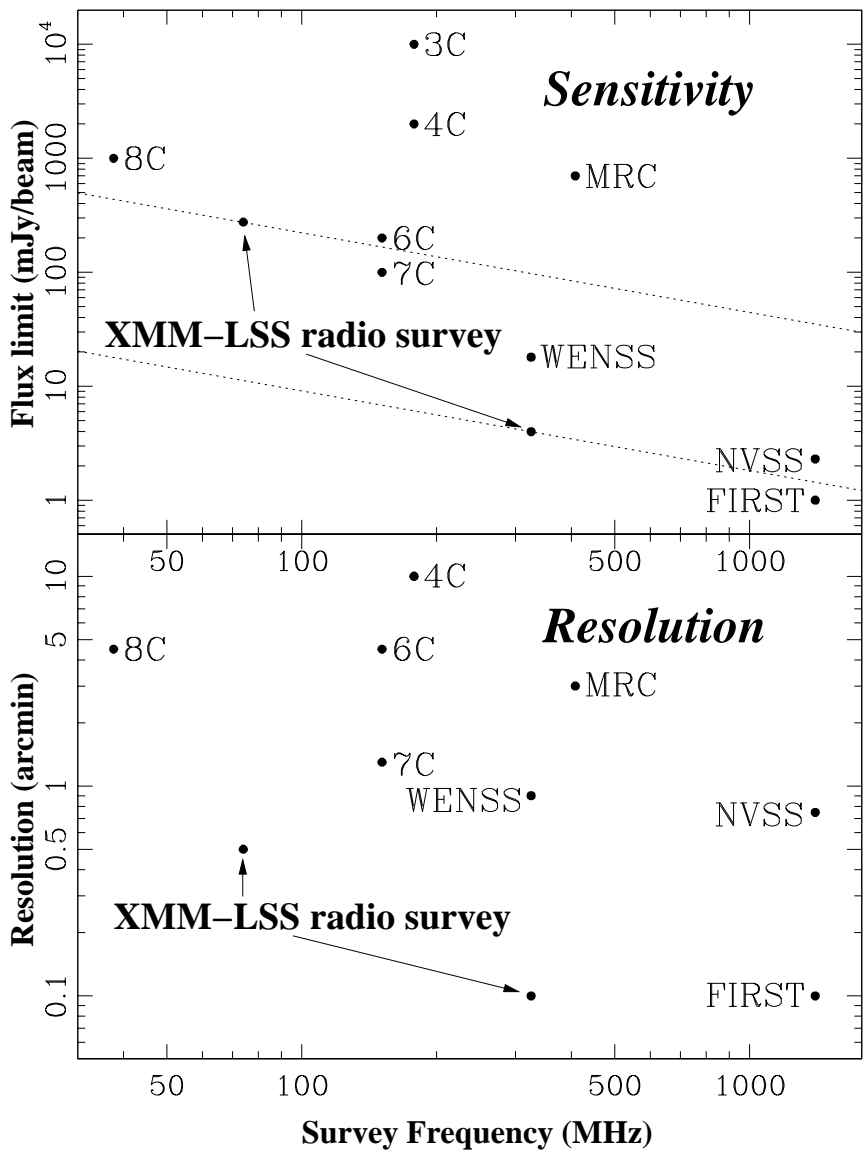}
\caption{The sensitivity and resolution of our XMM-LSS low frequency 
radio survey compared to that of the major low frequency radio surveys.
The dotted lines on the sensitivity plot (top) are lines of ``constant''
limiting flux assuming a spectral index of $\alpha = -0.7$. \label{survey.fig}}
\end{figure}

Though no published surveys exist at 74 MHz, we can compare our sensitivity 
to surveys at 38 MHz and 151 MHz by assuming a typical spectral index.  If 
we apply a spectral index of $\alpha = -0.7 (S \propto \nu^{\alpha})$, our 
74 MHz limiting flux density of 275 mJy/beam becomes 170 mJy/beam at 151 MHz
and 440 mJy/beam at 38 MHz.  Using this standard, our 74 MHz observation is 
about 
twice as deep as the 38 MHz 8C survey \citep{1990MNRAS.244..233R}, about 
20$\%$ more sensitive than the 151 MHz 6C survey \citep{1988MNRAS.234..919H} 
but only about 60$\%$ as sensitive as the 151 MHz 7C survey 
\citep{1998MNRAS.294..607V}.  Our resolution is about an order of magnitude
better than the 6C and 8C surveys and about a factor of 3 better than the
7C survey.  Figure \ref{survey.fig} shows the sensitivity and resolution of 
our XMM-LSS radio surveys at both 74 MHz and 325 MHz in comparison to major 
low frequency radio surveys.

As figure \ref{survey.fig} shows, at both frequencies, our survey has a 
unique combination of both depth and resolution.  The high resolution was 
desirable to help classify radio sources (eg. compact versus double lobe), 
to reduce classical confusion, and to allow better optical 
identification.  Using customized source-finding software, we identified 
and characterized 211 sources at 74 MHz and 256 sources at 325 MHz, many 
of which are resolved doubles.

This paper will be organized as follows:  Section \ref{obs.sec} will
describe our observation strategy, and Section \ref{red.sec} will 
explain in detail our data reduction procedure.  Our source finding
method will be covered in Section \ref{find.sec}.  That will be followed
by a description of the results (Section \ref{res.sec}), error analysis
(Section \ref{err.sec}) and survey completeness (Section \ref{com.sec}).
We end with our conclusions and future plans in Section \ref{con.sec}.

\section{Observations}
\label{obs.sec}

\subsection{Field Selection}
For the XMM-LSS survey, a field location was chosen centered at 
RA = $2^h 16^m$ and $\delta = -5^\circ$ mainly because of its high 
galactic latitude ($b = -58^\circ$), low $N_H$ and high XMM visibility 
(about $20\%$).  An additional consideration was easy visibility from 
both the northern and southern hemispheres for follow-up from 
globally dispersed observatories.

At 325 MHz, the primary beam is 2.68$^{\circ}$ (FWHP) in diameter which 
can only cover about 6\% of the XMM-LSS region.  Therefore, we pointed 
at the location which would best cover the area guaranteed to be imaged 
in the first round of XMM-LSS observations.  Our pointing center was at 
RA = $2^h 24^m 40.0^s$ and $\delta = -4^\circ 30' 0.0''$.  At 74 MHz, 
the primary beam has an $11.9^{\circ}$ diameter (FWHP) and the entire 
XMM-LSS field is covered.  

\subsection{Observation Strategy}

The large fields of view involved caused us to consider the non-coplanar 
nature of the VLA for non-snapshot observations. 
The geometric distortion caused by the non-coplanar nature of
the VLA leads requires a 3-D, rather than 2-D, Fourier inversion
of the visibility data to produce sky maps. This is a
computer-intensive process, although it is tractable, with the
greatest computational burden arising from observations with the
target at low elevations. In order to reduce the computational burden we 
therefore chose to observe in two 4-hour runs centered at transit on two 
different days, rather than a single 8-hour run. We observed 
simultaneously at 325 MHz and 74 MHz with bandwidths of 6 MHz and 1.5 MHz 
respectively.  With these configurations, bandwidth smearing would prevent 
us from imaging the entire primary beam of the VLA at either frequency in 
continuum 
mode. In fact the field of view in which bandwidth smearing expands the 
width of sources by less than a factor of two is only $0.37^{\circ}$ and 
$1.4^{\circ}$ for 325 MHz and 74 MHz respectively, a small fraction of the 
primary beam diameter for each, $2.68^{\circ}$ and $11.9^{\circ}$ 
respectively.  Therefore, we observed in spectroscopic mode to avoid 
problems of bandwidth smearing and to allow more efficient removal of 
radio-frequency interference (RFI).
At 74 MHz, the central frequency was 73.8 MHz, and we used 128 
channels, each of width 12.2 kHz.  At 325 MHz, we used 32 channels each of 
width 193.3 kHz.  We centered the 325 MHz observations for each 4-hour run 
at slightly different frequencies, 321.5 MHz and 328.5 MHz, in order to 
improve the combined $uv$ coverage.  We observed with an integration time 
of $\Delta t = 6.6$ seconds, the shortest possible with our frequency 
configuration.  This caused the predicted reduction in peak flux from 
time-smearing at the primary beam half power point to drop from $4\%$ (with 
the standard $\Delta t = 10s$) to $1.8\%$ (with $\Delta t = 6.6s$).  Our 
total time on source was 5.3 hours.  Observations took place on January 5th 
(LST 0100 to 0500) and January 12th (LST 0030 to 0430) of 2001.   

\section{Data Reduction}
\label{red.sec}

\subsection{Initial Calibration and RFI Removal}

At both 74 MHz and 325 MHz, we first performed a bandpass calibration on 
Cygnus A.  As Cygnus A is resolved, we referenced the calibration to 
models taken from previous observations at each frequency, scaled to 
match the absolute flux density measurements of \citet{1977A&A....61...99B}.  
We then performed an amplitude and phase calibration.  For the 325 MHz data, 
we used the flux and phase calibrators 3C48 and J0204+152, respectively, 
while at 74 MHz we used Cygnus A for both flux and initial phase 
calibration.

After calibration we focused on RFI removal.  For this, 
we used the AIPS task FLGIT, in which one can set overall clipping levels 
for the spectral data as well as remove data which fall too far off a 
linear spectral fit.  In practice, this removes most of the RFI 
which is strong enough to noticeably degrade the final image.  We 
experimented with the FLGIT parameters to find an acceptable balance 
between removing all the clearly bad data and keeping most of the 
unaffected data.  For the 325 MHz data, we clipped every baseline visibility 
($\Delta t = 6.6s$, $\Delta \nu = 193.3 kHz$) above 20 Jy.  For the 
74 MHz data, we clipped all visibilities 
($\Delta t = 6.6s$, $\Delta \nu = 12.2 kHz$) that were either above 
1,200 Jy or more than 500 Jy off the spectral fit.  The predicted noise
levels per visibility per channel are approximately 140 Jy and 2.8 Jy for 
74 and 325 MHz respectively.  As these clip levels are well above the 
total flux in each field as well as the noise levels, we are certain to 
have only removed corrupted data.  This removed $15.5\%$ and $4.0\%$ of 
the 74 MHz and 325 MHz data respectively.  

\subsection{325 MHz Data}

At 325 MHz, we removed three channels from each end of the bandwidth at 
both frequency settings, where
the gains are too low to achieve reliable calibration.  We were left with 
25 channels, each of width 193.3 kHz.  No channel averaging was done because 
that would have introduced 
noticeable bandwidth smearing toward the edges of the field, already 
at $4\%$ without any averaging.  For each baseline, each channel is
counted as a separate visibility.  Final flagging 
was done by hand with the AIPS task TVFLG in order to remove any remaining 
data which 
appeared corrupted, resulting in the removal of less than $1\%$ of the 
remaining data.  

The data from all channels were combined to produce images.  As the effects 
of a non-coplanar array become unmanageable in a field this large, the 
entire primary beam cannot be imaged at once with a standard 2D Fourier 
inversion. Instead a pseudo-3D treatment \citep{1999sira.conf..383P} was 
implemented in which we divided the primary beam area of diameter 
$2.68^\circ$ into 439 smaller fields (facets) each 
$350\times350$ pixels with a $1.5''$ pixel spacing.  For each facet, the 
phase center is reset to the center of that facet, and the synthesized 
beam used for deconvolution is recalculated for that location.  When 
combined, the facets cover the entire primary beam area in a single map 
6000 pixels across.  
Since bright sources outside the primary beam can cause sidelobe confusion 
in this map area, we also imaged an additional 49 facets outside the primary 
beam which contained bright NVSS sources, so that their sidelobes could be 
cleaned from the region of interest.  Deconvolution was performed with 
the CLEAN algorithm (\citet{1974A&AS...15..417H}, \citet{1980A&A....89..377C}).

Before imaging, phase only self-calibration was done according to a model 
of the source distribution in the primary beam predicted by applying a 
spectral index of $\alpha_{325}^{1400} = -0.75$ to the NVSS catalog.  The
AIPS task FACES was used to create this model.  Though assuming a constant
spectral index is not completely realistic, in practice this method generally 
produces adequate phase calibration for an initial image.  A further 
advantage to this method is that it removes any overall position shift in 
the field due to the ionosphere, which is generally of order $\sim10''$ 
at 325 MHz.  Later,
we performed two further iterations of phase-only self-calibration with 
the actual 325 MHz image, for further refinement.  The resulting image had 
an off-source RMS noise of 0.8 mJy/beam at the field center.  Since the 
thermal noise is predicted to be 0.3 mJy/beam and classical confusion noise
is negligible at this resolution, the sensitivity of the image is most likely 
sidelobe-confusion limited.  We corrected the primary beam attenuation 
away from the field center by dividing by the primary beam pattern with 
the AIPS task PBCOR.  This caused the noise level to increase toward the 
edges of the primary beam region.  The synthesized beam size was 
$6.3''\times5.3''$, and the image was convolved to a circular $6.3''$
resolution for source finding.

\subsection{74 MHz Data}

At 74 MHz, as for 325 MHz, we removed channels from the edges of the bandwidth 
which appeared to be of poor quality.  We were left with 120 channels, of 
which every 5 were averaged to produce 24 channels, each of width 61.0 kHz.  
We did not average any further as the bandwidth smearing at the field
edge for this channel width is already $7\%$.  Final flagging, in addition 
to the previous use of FLGIT on the initial spectral data, was done by 
hand with TVFLG in order to remove any remaining data which appeared 
corrupted, resulting in the removal of less than $1\%$ of the remaining data.

Checking the field beforehand against the NVSS database showed there to 
be one dominant source in this field, 3C63 
(source J0220.9$-$0156 in Tables \ref{table3} and \ref{table4}).  This 
source's flux density
was previously measured to be 34.0 Jy at 80 MHz \citep{1990PKS...C......0W}.  
Preliminary phase-only self calibration was performed by mapping only
this source.  We then included other sources in the field into 
the model, anticipating that this would improve the self-calibration as it 
had for the 325 MHz data.  This turned out not to be the case for 74 MHz. 
The assumption fails because of the breakdown at this frequency of the basic 
assumption 
of conventional self-calibration, that the ionospheric phase distortions 
over the field of view can be characterized by one time variable number 
per antenna.  This is because lines of sight through the ionosphere to 
sources on opposite sides of the large field of 
view are separated by large linear distances over which the ionosphere's
properties differ.  This radio-frequency ``seeing'' limitation is referred 
to as the breakdown of the ``infinite isoplanatic patch assumption'' 
and angle independent self-calibration is not applicable.

\subsubsection{The Ionosphere at 74 MHz}

\begin{figure}
\epsscale{1.00}
\plotone{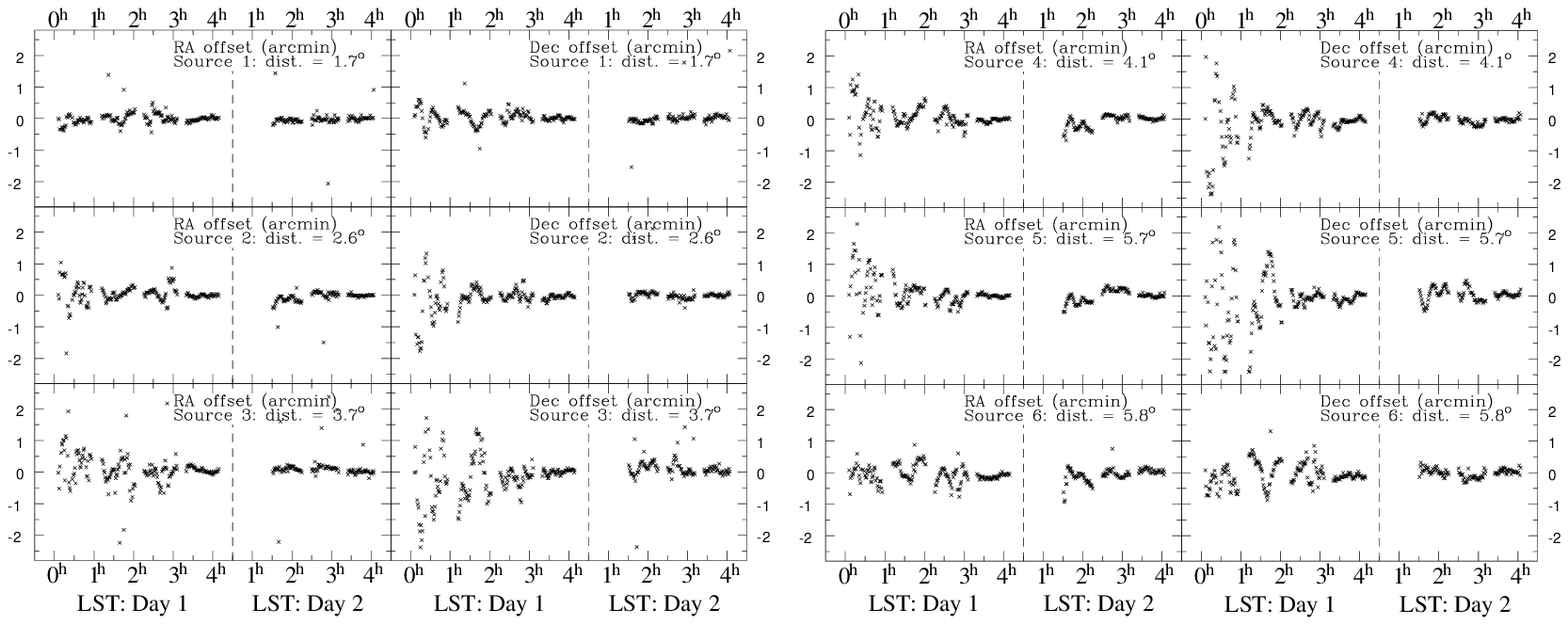}
\caption{The movement of various sources in the primary beam relative
to 3C63.  The distance of each source from 3C63 is labeled.  It seems
clear that the closest source, at a distance of $1.7^\circ$, moves less
with respect to 3C63 than do sources farther away.  The ''seeing'' due to 
the ionosphere appears to vary greatly in time as well.
\label{wander.fig}
}
\end{figure}

The ionosphere, which has a significant but manageable effect at centimeter
wavelengths, causes such large phase distortions at low frequencies 
($\nu < 150$ MHz), that this region of the electromagnetic spectrum has 
remained unexplored at sub-arcminute resolution until the 
relatively recent onset of the 74 MHz VLA 
(\citet{1991ritt.proc..249K}; \citet{1993AJ....106.2218K}).
Because of this and the fact that the ionosphere is the primary issue in 
74 MHz data reduction, we will explain some of the main issues involved 
in this section.

A low frequency interferometer is extremely sensitive to {\em differences}
in the total electron content (TEC) above pairs of antennas. For our 
discussion of the effects of the ionosphere on the 74 MHz data, the 
ionosphere can be approximated by a thin screen phase
delay model which varies in both space and time across the field of 
view.  At the height of the ionosphere (which varies with the time of 
day) the $11.9^{\circ}$ field of view at 74 MHz corresponds to a 
patch of sky roughly 100-200 km in diameter.  
Very large scale variations ($\geq1000$km) in the TEC impose a linear phase 
gradient across
the field of view which causes all sources to shift, in
unison, as a function of time.  This time-varying position wander causes
smearing of sources, lowering their peak flux.  Somewhat 
smaller ionospheric structure ($\geq100$km) produces phase gradients that 
are not 
constant across the field of view.  This introduces differential
refractive effects, in which pairs of sources no longer wander in unison.
Finally, much smaller scale structures (smaller than the size of the array,
or $\geq10$km) generate source distortions which are no longer simple phase 
shifts. These 
combined effects can cause individual source peaks to be lowered below the 
noise level and remain
undetected. The effects are somewhat analogous to the ``seeing'' effect in
optical telescopes.  At 74 MHz the refractive effects can shift objects 
by more than an arcminute on a scale
of tens of minutes (and just a few minutes during particularly 
bad periods).  With our A-array resolution of $30''$, this is a serious 
problem.  

Phase-calibration can solve certain aspects of this problem by removing 
the first order refractive effect.  A model 
of the field can be made from a very short snapshot observation.  The 
time interval of the snapshot should be a time period too short for 
the source-wander to be significant, generally about one minute.  Phase-only 
self-calibration of all the data to this model will align the source 
positions at all times to that of the snapshot.  Though there will still 
be a source location offset, it will be constant in time removing the 
smear and restoring the original morphology and peak flux.  

\begin{figure}
\epsscale{0.80}
\plotone{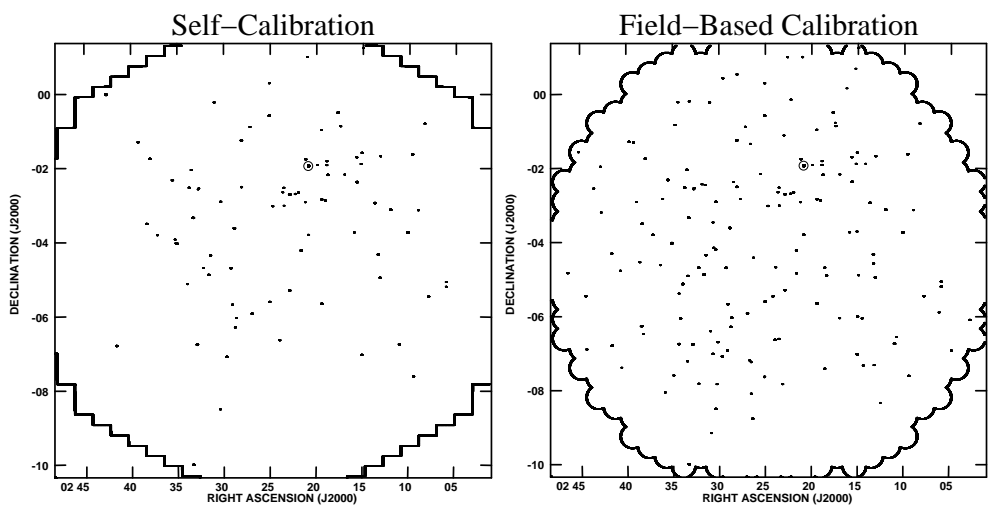}
\caption{Images of the 74 MHz primary beam area produced using 
self-calibration (left) and VLAFM (right).  The bright source 3C63 
is circled.  This is a contour map in which the first contour is set
so far above the noise, 400 mJy/beam, that only actual sources appear.
Since the synthesized beam is so small compared to the entire field,
the sources simply appear as dots at this scale. 
\label{4A.maps.fig}}
\end{figure}

The limitation of this technique is that the ionospheric phase offsets 
vary across the $11.9^\circ$ primary beam, producing different position 
shifts in different regions.  However, standard self-calibration does 
not compensate for the higher order refractive effects that cause 
differential refraction and higher order source distortions.  This is why 
adding sources other than 3C63 to our self-calibration model did not 
improve the images.  Though the position of 3C63 was fixed, the positions 
of other sources in the field were not.  Figure \ref{wander.fig} shows 
the position offsets as a function of time for six bright sources in the 
field.  These position offsets are measured relative to the expected 
position of the sources relative to 3C63.  This plot shows that the 
source locations ``wander'' over the course of the observation, sometimes 
by more than $2'$.  
Table \ref{table1} shows the RMS offsets for each source in 
Figure \ref{wander.fig} during two time intervals.  The first time 
interval is the first 200 minutes of the observation during which
the ionosphere was very active.  The second time interval consists of 
the remainder of the observation, when the ionosphere was relatively
``quiet''.  The average offsets are drastically different for each
time interval.  It is also clear that in most cases, the average 
offsets are greater in sources which are farther away from 3C63.  
This makes sense as one would 
expect the applicability of the 3C63-based calibration to decrease with 
increasing distance from 3C63.  This is also seen in an image produced 
from the self-calibrated data set of the entire primary beam 
(Figure \ref{4A.maps.fig}, left).  Here, the density of sources visibly 
decreases with increasing distance from 3C63 as the ionospheric smearing 
causes the source peaks to decrease.  (Primary beam attenuation causes the 
source density to fall off toward the field edges as well.)

Recently, J. Condon and W.~D.~Cotton of NRAO developed an imaging algorithm
to apply the position-dependent calibration needed to correct for the higher
order refractive effects across the full primary beam \citep{2002URSI....C}. 
Implemented in the AIPS task VLAFM (VLA-Four-Meter), it produces images of
facets placed at the locations of the brightest sources known from the 
NVSS catalog to be in the field.  For each observation time interval 
the position of the source with respect to its known position is 
measured.  The time interval is chosen to be small enough that the source
wander over that time is negligible, which in our experience is generally
one or two minutes.  Based on the source shifts as a function of position 
in the sky, a $2^{nd}$ order Zernike polynomial model of the ionosphere 
is produced for each time interval.  Typically not many more than about 6 
sources in a given field are bright enough for their positions to be 
accurately measured in such a short time interval with the VLA, and it is 
this that sets the limit to the order of Zernike polynomial that VLAFM was 
designed to fit.  When mapping the field, this model of the ionosphere is 
used to apply a different phase calibration to each facet of the image.  
Deconvolution is performed by subtracting the clean components from the 
$uv$ data, after reversing the ionospheric corrections, and remapping
the field with the ionospheric corrections restored.

We applied VLAFM to our self-calibrated data set with striking results 
in the resulting image (Figure \ref{4A.maps.fig}).  With 
position-dependent phase calibration, the source density becomes uniform 
across the field, with many more sources detected.  It is clear that 
VLAFM greatly improved the ionospheric distortions in the image.  
Distortions from the smallest scale ($<$ 10 km) ionospheric structures demand
higher order Zernike terms or alternative algorithms. Ignoring them
ultimately limits our image fidelity; efforts to develop solutions to this
challenge are ongoing.  

\subsubsection{Final 74 MHz Image}

We use the image produced with VLAFM as our final image.  We used a 
circular restoring beam with FWHM = $30''$, and the pixel 
spacing was $7.5''$.  We set the facet size to 360 pixels and we 
needed 349 facets to cover the primary beam area.  Another 11 facets
outside of the primary beam area were included because the NVSS
catalog indicated they contained bright outlier sources.  With the facets 
combined, the entire field is 5600 pixels on a side.  As with 
the 325 MHz image, we corrected the primary beam attenuation away from 
the field center by dividing by the primary beam pattern with the AIPS 
task PBCOR.  This resulted in somewhat higher noise toward the field 
edges.  

\section{Source Extraction}
\label{find.sec}

With millions of pixels and hundreds of sources, we needed an automated
and consistent method for locating and characterizing sources in the maps.
The algorithm we used is the AIPS task VSAD, which was produced for use in 
the NVSS survey \citep{1998AJ....115.1693C}.  
This task searches the field for islands of emission with peak flux
density greater than some specified cutoff level, and then fits
the island with a model composed of multiple Gaussian components,
with each parameterized by 
central position, peak flux density, major and minor axes and 
position angle.  

To distinguish ``real'' sources from noise, we first determined the root mean 
square noise ($\sigma_{rms}$) of the image.  This was measured for the central 
region of each image with the AIPS task IMEAN.  We then set the peak flux 
cutoff in VSAD at the 5$\sigma_{rms}$ level.  Additionally, we required that 
the {\em integrated} flux of each source, calculated as the peak flux 
density per beam times the Gaussian area in units of beamsize, also
be greater than 5$\sigma_{rms}$.  Sidelobes from very bright sources could 
be confused as sources.  Therefore, we removed sources with peak flux less 
than 0.04 times that of a nearby source.   
Smearing effects, such as from bandwidth and time averaging as well as 
from inaccurately removed ionospheric refraction, can lower the peak flux 
density of a source.  However, if the ionospheric effects are 
primarily refractive this flux is not lost, but merely spread out. 
Therefore, we regard the integrated flux, calculated as the peak flux 
density per beam times the Gaussian area in beams, as the more robust flux 
density measurement, particularly at 74 MHz, and we use this for the 
remainder of the paper.  

\section{Results}
\label{res.sec}
\subsection{325 MHz Results}

\begin{figure}
\epsscale{0.80}
\plotone{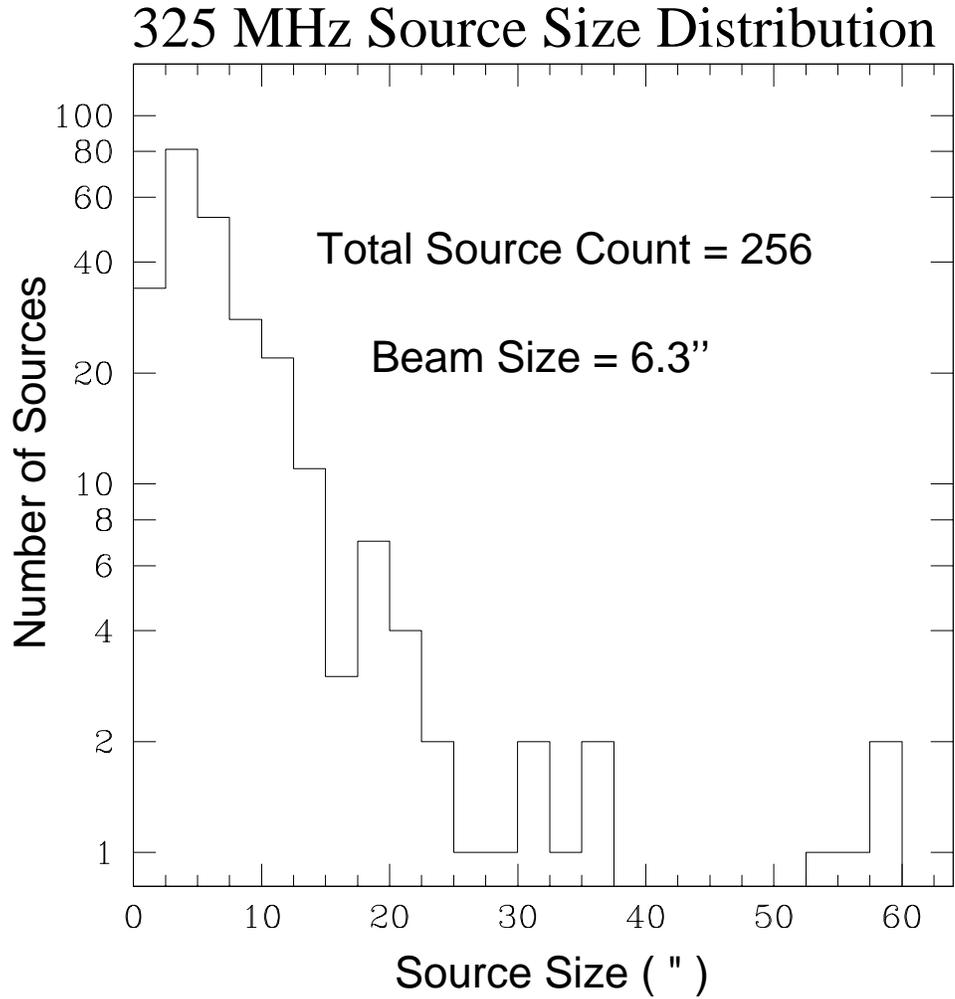}
\caption{Histograms of the deconvolved source size distribution for the 
325 MHz survey.  
Source size plotted is either the major axis of the fitted Gaussian for 
the single sources or the largest source separation for the multiple 
sources.
\label{size.fig}}
\end{figure}

\begin{figure}
\epsscale{0.80}
\plotone{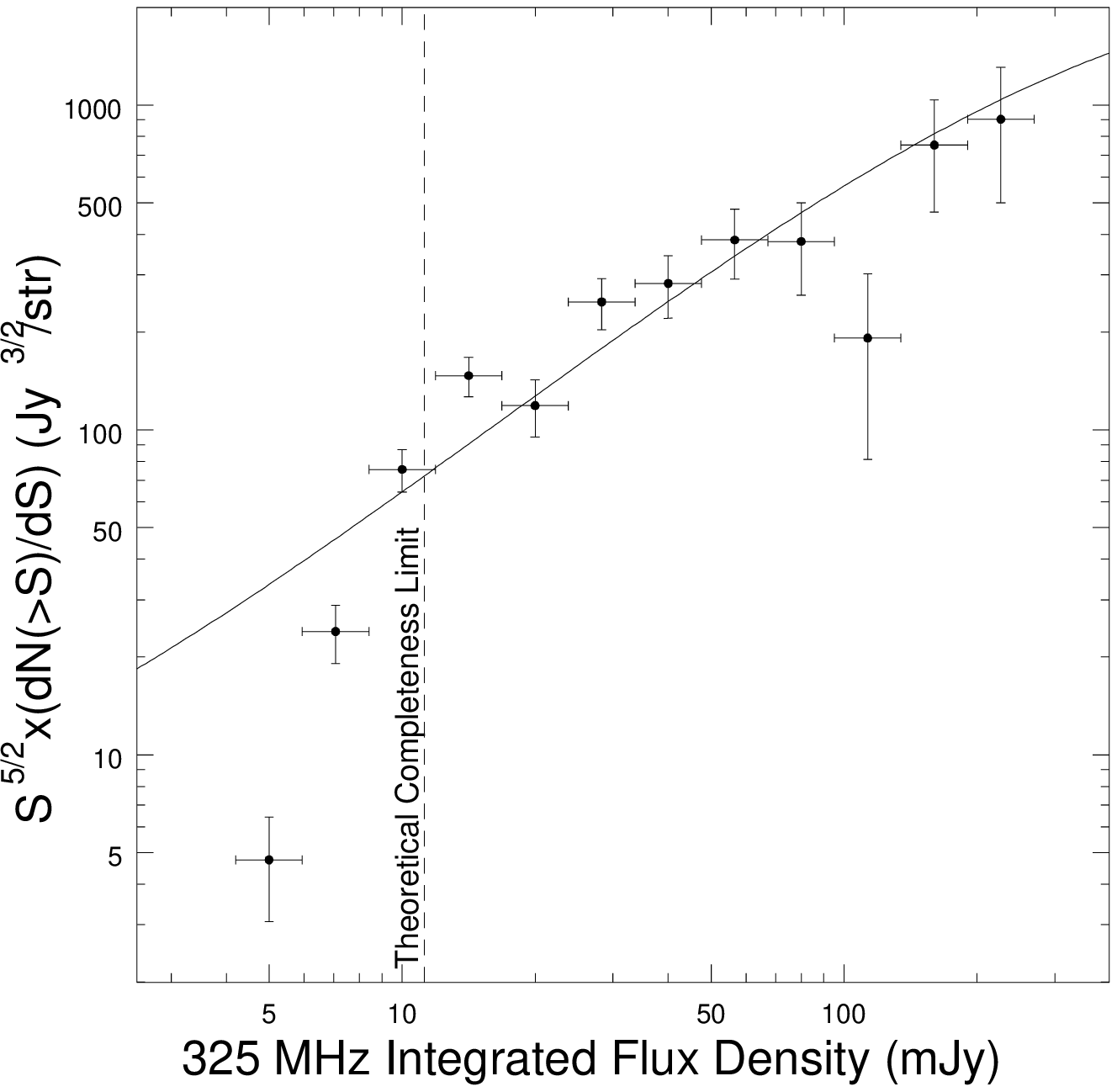}
\vspace{0.0in}
\caption{Euclidean normalized differential source counts for the 
325 MHz image.  Also plotted is the curve showing the source 
counts from a deep WSRT survey \citep{1991PhDT.......241W}.
\label{lnlsp.fig}}
\end{figure}

\begin{figure}
\epsscale{0.90}
\plotone{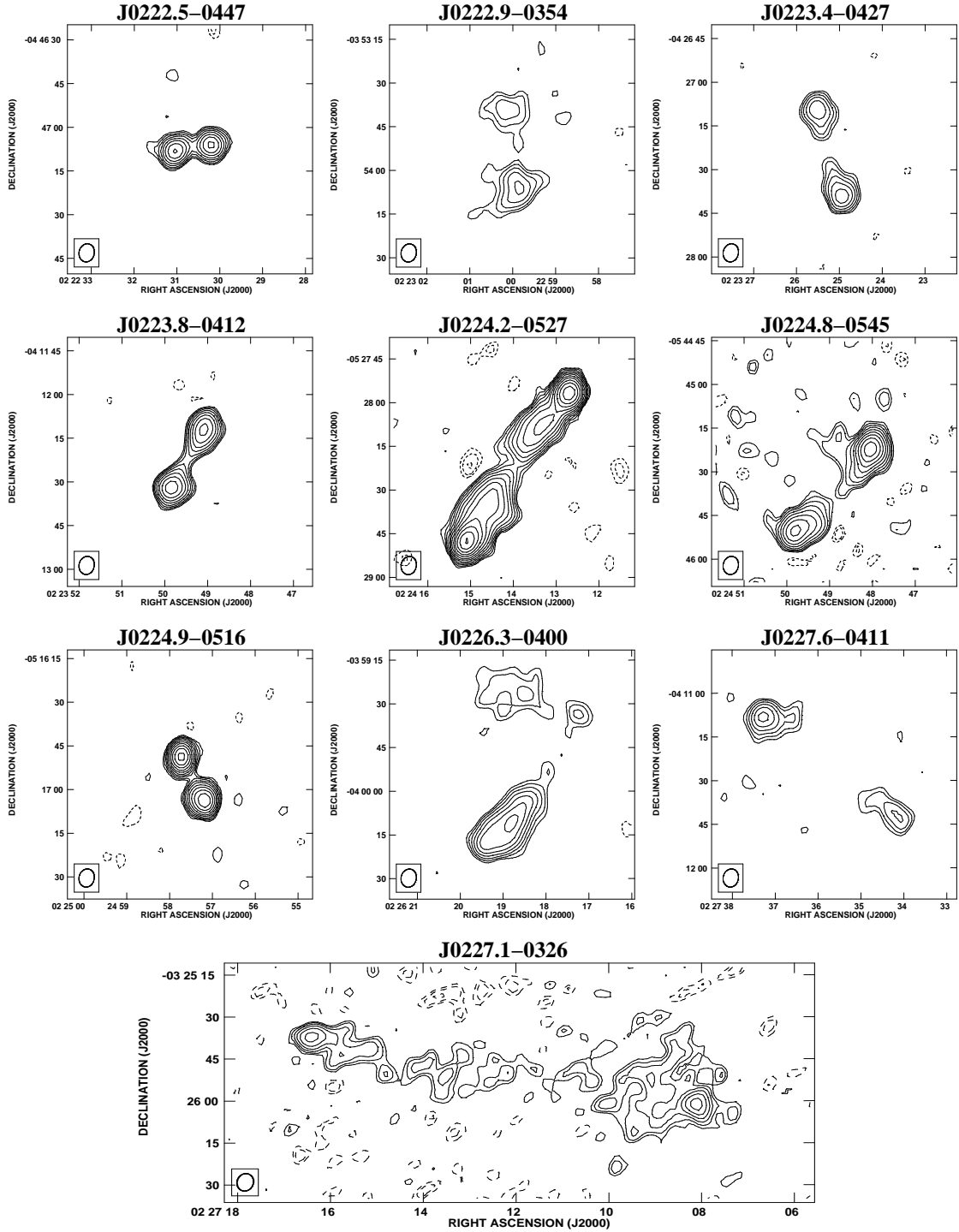}
\caption{A sample of some of the larger resolved sources from the 
325 MHz observation.  The angular scale is the same in each image.  
Contour levels are at 2.5 mJy/beam $\times$ 
(-2, -1.4, -1, 1, 1.4, 2, 2.8, 4, 5.8, 8, ...) for each image.
\label{psources.fig}}
\end{figure}

With the criteria of Section \ref{find.sec}, we detect 256 sources at 
325 MHz.  Of these, 219 were single (``S'') sources and 37 were multiple 
(``M'') sources.  (Multiple sources were defined as any sources that were 
closer together than $60''$.)  We classify as ``unresolved'' those sources 
in which the 
major axis of the Gaussian fit was indistinguishable from the beam size at
the $2\sigma$ level.  Of the 219 single component sources, 116 were 
unresolved.  A size distribution histogram for all 256 sources is shown 
in Figure \ref{size.fig}.  Images of several of the larger resolved sources
are shown in Figure \ref{psources.fig}, and we present our complete catalog 
in Table \ref{table2}.  With this many sources, some statistics about radio 
sources at this frequency and flux density limit can be determined.  

In Figure \ref{lnlsp.fig}, we show the Euclidean normalized differential 
source counts.  The source density is calculated as the number of sources 
detected in each flux bin, divided by the solid angle of the full field 
of view.  The completeness level shown is that for the field edge, where
the noise is highest.  For this plot, we are limited on the low-flux 
density side 
by the flux density limit of our observation, and on the high-flux density
side by the relatively small area of our survey.  Therefore, compared to 
the WENSS measurement of the differential source count 
(Rengelink, et al. 1997), which had a much larger survey region but 
higher limiting flux density, we measured a very different 
region of this curve, though there is some overlap.  The curve on the 
plot is a polynomial fit to the differential source count from a deep 
Westerbork survey \citep{1991PhDT.......241W}.  In general, we find close 
agreement between their results and our own.  However, there is possibly 
a trend with our source counts being slightly higher at lower flux 
densities, and slightly lower at higher flux densities.  If real, this 
could possibly be explained by the fact that our resolution was nearly 
an order of magnitude higher.  Therefore, some of the extended emission
from the bright sources could have been resolved out of our observations.
This would result in some bright sources at low resolution being counted
as faint sources at high resolution.

\begin{figure}
\epsscale{0.80}
\plotone{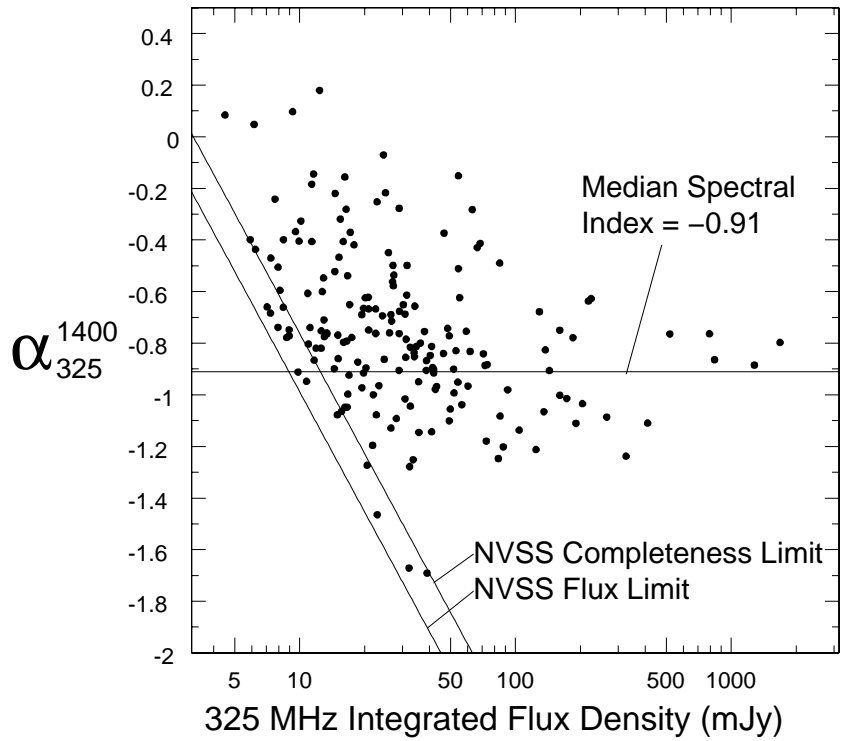}
\caption{Spectral indices ($\alpha$ where $S \propto \nu^{\alpha}$) 
from comparing the 325 MHz integrated flux density to the NVSS flux 
density plotted versus the 325 MHz integrated flux density.  The 
limiting flux density and completeness level of NVSS restrict the 
parameter space somewhat.  Because the 325 MHz image has much greater 
resolution, the spectral index could be overestimated for resolved 
sources.
\label{pspec.fig}}
\end{figure}

We compared our source list to the NVSS catalog, searching for 
counterpart sources within $45''$ of each of our sources.
We found NVSS counterparts to 181 out of 256 sources.  The 
remaining 75 sources most likely have 1.4 GHz fluxes below the 
NVSS sensitivity limit of 2.5 mJy/beam, and we can calculate the
upper limits to their spectral indices.  Most of the spectral index
upper limits seems quite reasonable, although a handful require quite
steep spectra ranging down to as low as $\alpha = -1.56$.
In Figure \ref{pspec.fig} we plot the spectral indices 
($\alpha_{325}^{1400}$) of all the sources with counterparts in NVSS.  
The difference in resolution ($\sim6''$ in our map, $45''$ for NVSS) between 
our data and NVSS means that for extended sources, we may miss some of the
325 MHz flux and so estimate the spectral index to be flatter than it truly 
is.  The median 
spectral index was $\alpha_{325}^{1400} = -0.87$ (this includes upper 
limits for sources with no NVSS counterpart).  There are six sources
with inverted spectra ($\alpha_{325}^{1400} > 0$), though we 
would need matched resolution to verify this.  Seven steep-spectrum 
sources ($\alpha_{325}^{1400} < -1.3$) were revealed, three with measured 
spectra and four with spectral limits.  As our spectral 
index estimate is, if inaccurate, certainly an overestimate of $\alpha$, we 
are confident that these are indeed steep-spectrum objects.  

\subsection{74 MHz Results}

With the criteria of Section \ref{find.sec}, we detect 211 sources 
at 74 MHz.  Of these, 196 were single (``S'') sources and 15 were multiple 
(``M'') sources.  (Multiple sources were defined as any sources that were 
closer together than $60''$.)  As with the 325 MHz data, we classify as 
``unresolved'' 
those sources in which the major axis of the Gaussian fit is 
indistinguishable from the beam size at
the $2\sigma$ level.  Of the 196 single sources, 170 were unresolved,
though as we discuss below, the size measurements at 74 MHz have very
high uncertainties.  
Figure \ref{4sources.fig} shows images of a sample of some of the larger 
single sources and multiple sources in the field.  We present our complete 
catalog in Table \ref{table3}.  

\begin{figure}
\epsscale{0.80}
\plotone{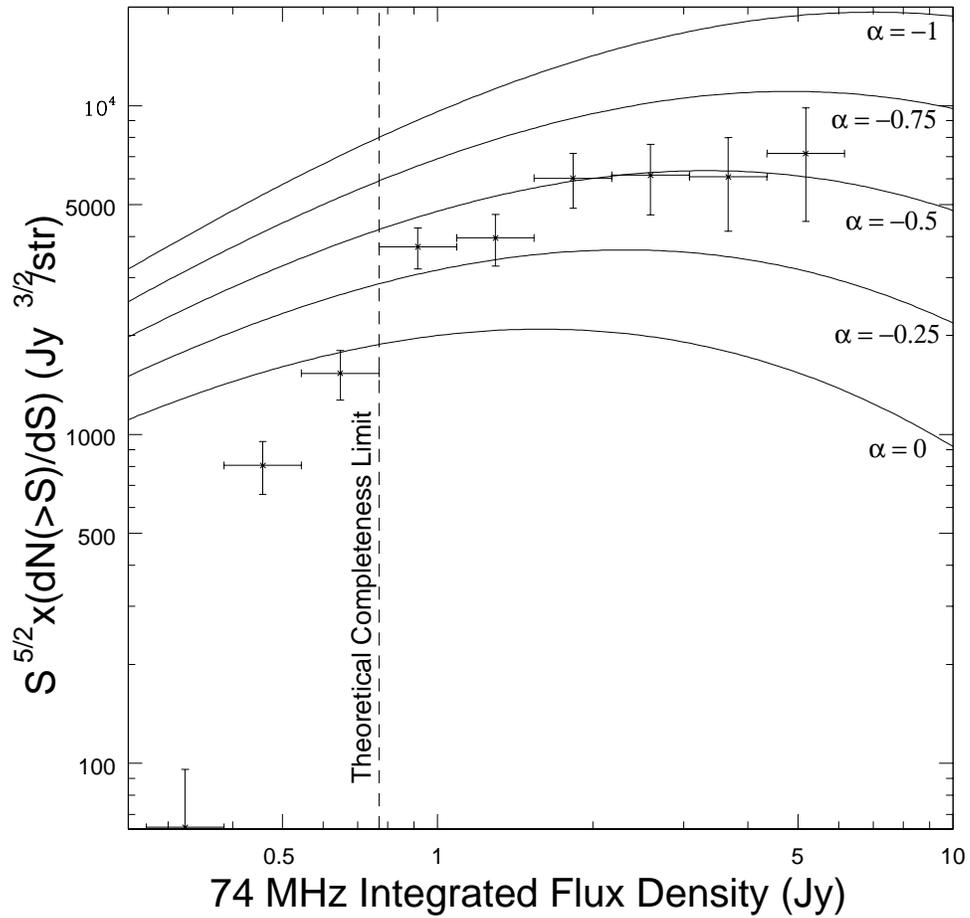}
\caption{Euclidean normalized differential source counts for the 
74 MHz image.  Also plotted are curves showing the source 
counts from a deep WSRT survey \citep{1991PhDT.......241W} scaled
to 74 MHz using various spectral indices as labeled.
\label{lnls4.fig}}
\end{figure}

We plot the Euclidean normalized source count for 74 MHz sources 
(Figure \ref{lnls4.fig}).  The completeness level shown is that for 
the field edge.  With no published surveys
at 74 MHz, there are no observationally determined source counts with 
which we can compare these results as we did for our 325 MHz source
counts.  However, one can adjust the 325 MHz source count curve to 74 MHz
by assuming a constant spectral index of sources.  This was done for 
several spectral indices ranging from 0 to $-1$, and these curves are also 
plotted on Figure \ref{lnls4.fig}.  Comparing these curves to our measured 
source counts shows that a relatively flat average spectral index of 
$\alpha_{74}^{325} \sim -0.5$ seems to be most consistent with the data.
We caution however, that source counts can be misleading in determining
the average spectral index because some fraction of sources become 
completely absorbed between 325 MHz and 74 MHz.    

\begin{figure}
\epsscale{0.80}
\plotone{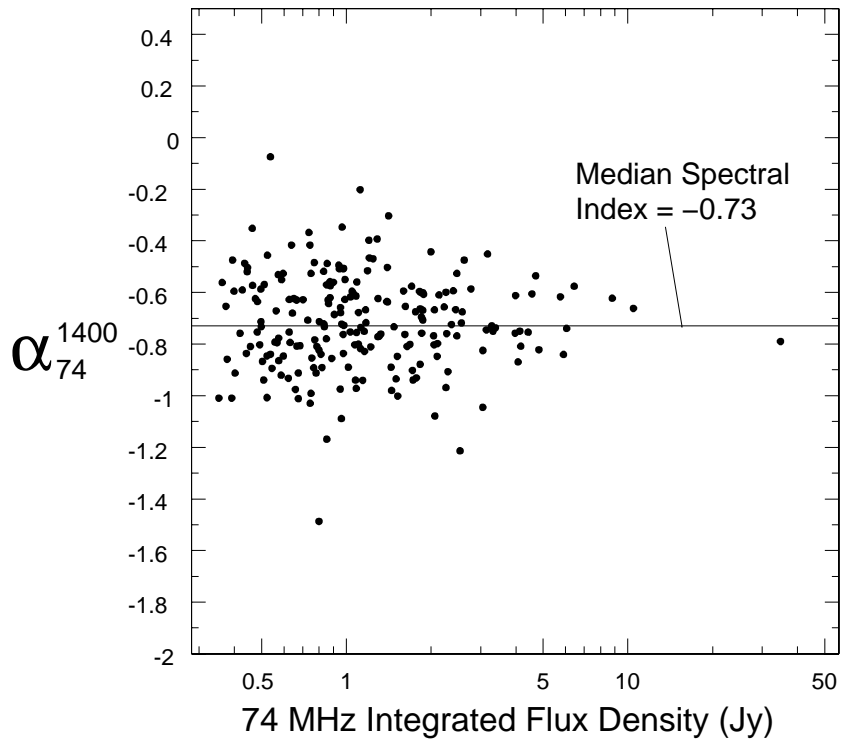}
\vspace{-0.5in}
\caption{Spectral indices ($\alpha$ where $S \propto \nu^{\alpha}$) 
from comparing the 74 MHz integrated flux density to the NVSS flux 
density plotted versus the 74 MHz integrated flux density.  The 
74 MHz resolution of $30''$ roughly matches the NVSS resolution of
$45''$.
\label{spec4.fig}}
\end{figure}

The 74 MHz resolution ($30''$) is a much closer match to that of 
NVSS ($45''$), and so we expect to derive accurate spectral indices
based on these comparisons.  Unlike at 325 MHz, the flux density 
limit at 74 MHz is so high that we expect all detectable sources 
to have NVSS counterparts.  In fact each 74 MHz source had an NVSS 
counterpart within $60''$.  We plot the spectral index
of all 74 MHz sources based on comparisons to the NVSS 
in Figure \ref{spec4.fig}.  We found a median spectral index
of $\alpha_{74}^{1400} = -0.72$.  This is significantly flatter 
than the median spectral index for the 325 MHz sources, indicating 
that in general, source spectra flatten at low frequencies.  This 
result is expected because of absorption effects as well as spectral
aging. 

Interestingly, in Figure \ref{spec4.fig}, 
we see no 74 MHz sources with inverted spectra as measured from 1.4 GHz, 
while we found six in our 325 MHz survey (Figure \ref{pspec.fig}).  
This seems likely to be a selection effect.  Assuming that inverted 
spectra occur because of synchrotron self-absorption, the flux is expected
to decrease dramatically ($S \propto \nu^{5/2}$) below the frequency at which 
self-absorption becomes significant.  Since our source population consists 
of objects which are relatively bright at 74 MHz, objects with inverted 
spectra, like gigahertz peaked 
sources, are simply too faint to be detected, and never enter the sample.  
The bright source population at 74 MHz is therefore dominated by larger 
objects like the lobes of 
radio galaxies which are not noticeably affected by synchrotron 
self-absorption.  

\subsection{Source Identifications from Literature}

\begin{figure}
\epsscale{0.90}
\plotone{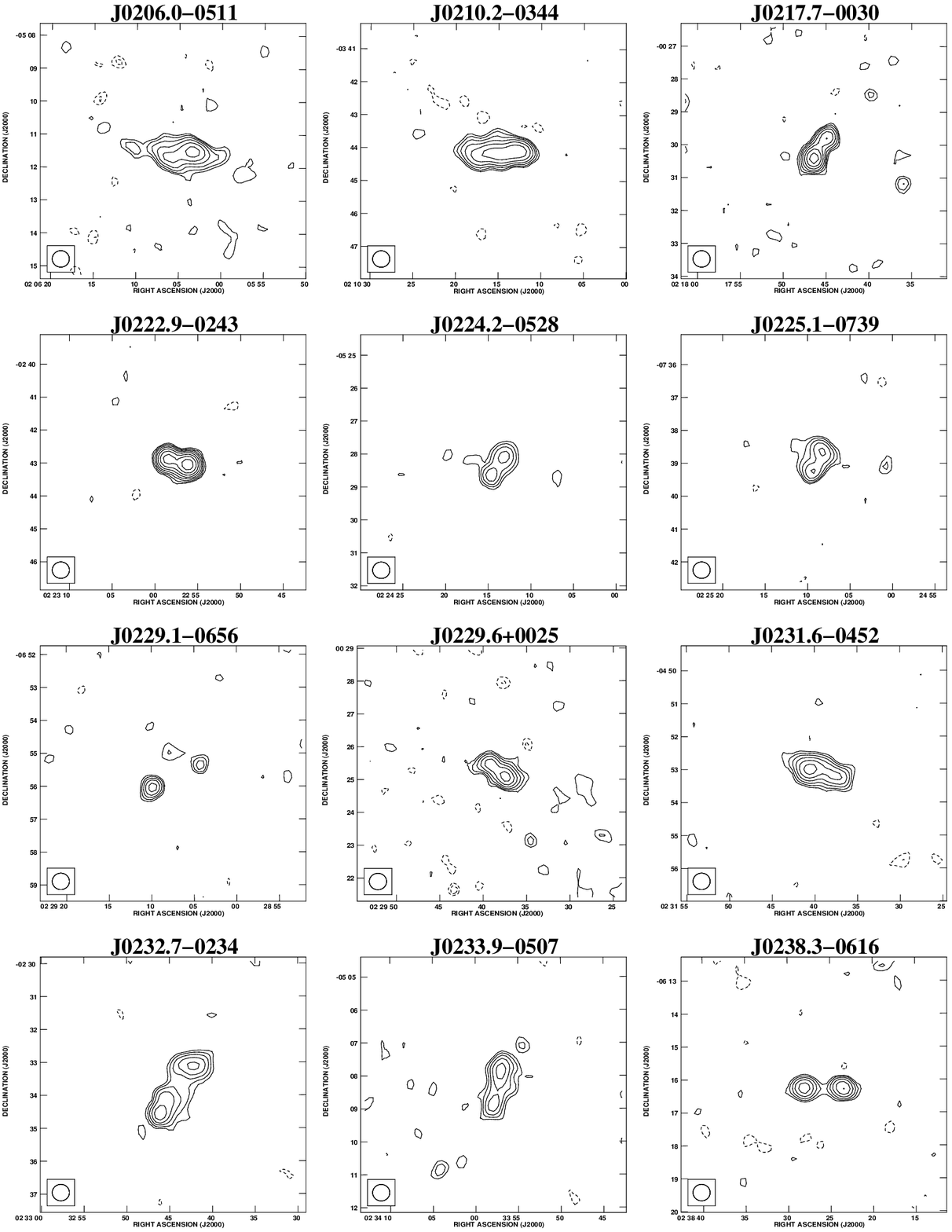}
\caption{A sample of some of the resolved sources from the 74 MHz 
observation.  The angular scale is the same in each image.  Contour levels 
are at 150 mJy/beam $\times$ (-2, -1.4, -1, 1, 1.4, 2, 2.8, 4, 5.8, 8, ...).
\label{4sources.fig}}
\end{figure}

Several sources we detected are described in the literature, many
of which have optical identifications and redshift measurements.
We made use of the NASA/IPAC Extragalactic Database 
(NED, \citet{1992NED11.R......1N})
to search for published descriptions of all the objects we detected
at both frequencies.  We classified as an identification any object
within $6''$ of our source detection, or within $3'$ for galaxy 
clusters.  In Table \ref{table4}, we list the 22 objects which have been 
identified optically as either a galaxy or QSO, and which also have 
had their redshifts determined.  Of these 22 objects, all were detected
at 74 MHz, yet only one, J0228.8-0337, was identified at 325 MHz as well.  
This is likely a selection effect, as our 74 MHz catalog consists of 
much brighter objects than our 325 MHz catalog.  We found that one object, 
J0238.3-0616, is only $1.6'$ from the known cluster Abell 362 
\citep{1989ApJS...70....1A}.  We detected this object at 74 MHz as
one of the largest extended objects at that frequency (seen in 
Figure \ref{4sources.fig}).  

One would expect radio sources detected at 74 MHz to have higher redshifts 
on average than those detected at higher frequencies.  This is because of 
the steep spectrum nature of high redshift radio galaxies, as well as the 
tendency for near-by compact radio sources to become synchrotron 
self-absorbed and not detected at 74 MHz.  The median redshift for these 
22 objects was $z = 0.8$.  This is roughly consistent with the median 
redshifts found in studies of sources at 1.4 GHz \citep{2001MNRAS.328..882W}.  
However, 18\% (4 out of 22) have redshift above $z > 2$, which is 
well above the roughly 9\% found by \citet{2001MNRAS.328..882W}.  Further,
these 22 objects with redshift in the literature do not constitute a 
complete sample, and could have a selection bias in favor of low redshift 
objects redshift.  Therefore, these findings seem consistent with 74 MHz 
sources having higher average redshifts than higher frequency source 
populations, though further study of a complete sample is necessary to 
confirm this.

\section{Accuracy of Results}
\label{err.sec}

The most relevant data we seek to extract from this survey are the 
positions and fluxes of all detectable sources.  We also present source
sizes, but as we discuss, these are compromised at 74 MHz by residual
ionospheric effects.  In this section we discuss the accuracy of these 
measurements.

The measurement error for each source parameter is a combination of two 
types of effects, one noise dependent and the other noise independent.
The noise dependent error relates to source fitting errors due to map 
noise, and these naturally decrease with increasing signal to noise ratio.
We calculate these according to the method described by 
\citet{1997PASP..109..166C}, equations 21, 41 and 42.
Noise independent errors are independent of the signal to noise ratio, and 
include calibration errors and, for 74 MHz, residual ionospheric errors.
These errors are discussed in the following sections.  

\begin{figure}
\epsscale{0.95}
\plotone{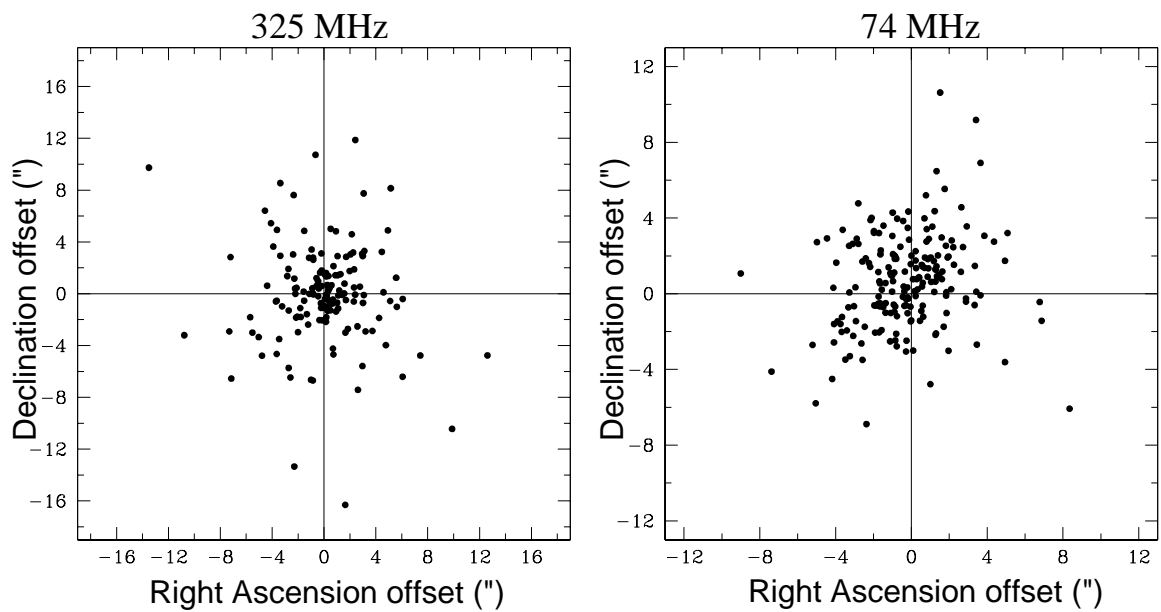}
\caption{Plots of position offsets with respect to NVSS counterparts.  
All single sources with NVSS counterparts are plotted: 153 and 196 
sources for 325 MHz and 74 MHz respectively.  The 325 MHz offsets 
are dominated by NVSS position errors, while the 74 MHz offsets are
dominated by 74 MHz position offsets.  As the NVSS sources corresponding
to 325 MHz sources are much fainter than those corresponding to 74 MHz 
sources, the position errors are much larger.  
\label{pos.fig}}
\end{figure}

\subsection{Noise-Independent Position Errors}

For both 325 MHz and 74 MHz observations, the phase calibration is 
referenced to the NVSS catalog.  Therefore, one portion of the 
noise-independent error is the NVSS calibration uncertainty values 
of $C_\alpha = 0.45''$ and $C_\delta = 0.56''$ for the right ascension and
declination axes respectively \citep{1998AJ....115.1693C}. 

For the 74 MHz, observations, this is not the only component of the 
noise independent error.  Another component is the error introduced 
by incompletely removed ionospheric shifts.  As the ionosphere in the 
field of view was approximated by only a $2^{nd}$ order Zernike 
polynomial, there could be residual shifts to source positions.  This 
effect was examined by comparing source positions to the positions of 
NVSS counterparts (Figure \ref{pos.fig}).  Although the NVSS resolution 
of $45''$ is
somewhat worse than the $30''$ resolution of our 74 MHz image, virtually
all NVSS sources are detected at a signal-to-noise ratio at least an 
order of magnitude greater than at 74 MHz.  Therefore, we assume that 
the NVSS position errors are negligible compared to those of our image,
and the position offset of our source from its NVSS counterpart is very 
close to its actual position measurement error.
The rms values of the position offsets, $3.0''$ and $2.5''$ for right 
ascension and declination respectively, were higher than one would 
expect with no residual ionospheric shifts.  Adding to the 
position error estimates a constant ionospheric error of $1.7''$ and $1.4''$ 
for right ascension and declination respectively brought the estimated 
errors into agreement with those measured with respect to the NVSS.  We 
therefore add in quadrature these values to the noise based position 
errors in calculating the total position errors.  

\subsection{Noise-Independent Flux Density Errors}

The noise-independent component of the flux density errors consists 
of the flux density scale uncertainty.  The overall flux density scale
was set using Cygnus A for 74 MHz and 3C48 for 325 MHz, and this is known 
to be reliable to 5$\%$ accuracy for each frequency \citep{2002K}.  
Therefore an additional 5$\%$ error is added in quadrature to the 
map noise based source fitting error \citep{1997PASP..109..166C}.

\subsection{Noise-Independent Source Size Errors}

At 74 MHz, in addition to the source size uncertainties due to fitting 
errors, we also must consider residual ionospheric effects.  
Incompletely removed ionospheric shifts for each time interval combine 
to smear sources in the final image.  This is somewhat analogous to the 
``seeing'' in optical images, and will cause source sizes to be 
overestimated.  It is therefore desirable to measure the amount of 
``seeing'' in an image to remove its effect from source size measurements.
For this discussion, we quantify the seeing, $\theta_{seeing}$, as the 
FWHM of point source imaged with that level of seeing.  If we measure 
a fitted source size of $\theta_{fit}$, and call our beam size 
$\theta_{beam}$, then we can calculate the deconvolved source size, 
$\theta_{source}$ to be:

\begin{equation}
\label{size.eqn}
\theta_{source}^2 = \theta_{fit}^2 - \theta_{beam}^2 - \theta_{seeing}^2
\end{equation}  

Unfortunately, there is no straightforward way to directly measure the 
ionospheric ``seeing'' in an image.  This is because of the difficulty 
in separating seeing effects from actual source sizes.  One could consider 
using, when available, higher resolution images at higher frequencies to 
determine the actual source sizes to isolate the seeing effect.  However, 
this approach has problems.  First of all, sources may not have the 
same size at 74 MHz, as they do at higher frequencies.  For example, 
a core-lobe source with a flat spectrum core and a steep spectrum lobe
could look very different at 74 MHz and at 325 MHz or 1.4 GHz.  Further,
higher resolution images can resolve out exactly the diffuse emission that 
74 MHz is actually biased toward.  We therefore discuss two imperfect
ways to measure the seeing, and our attempt to make due with this.

The first method for measuring seeing is based on the quality of fitting
ionospheric corrections to source offsets during data reduction.  In
each time interval, the rms offsets of sources from the fitted ionospheric
corrections can be calculated.  For our 74 MHz data, this method estimates
a seeing effect with FWMH of $10.3''$.  This is likely an overestimate 
because this figure includes fitting errors and the actual offsets of 
source centroids between 1.4 GHz and 74 MHz.  

Another method is to look at the median source size for all sources 
in the field.  The median is used to reduce the effect of the small 
number of source which are truly large objects.  Taking all single 
sources, we find the median value for the major axis to be $37.9''$.  
Deconvolving the $30''$ restoring beam implies a seeing FWHM of $23.2''$.
This is also clearly an overestimate, as many sources can be expected 
to have actual source sizes in this range.  In fact, it is likely that 
our very low frequency biases us toward large objects.

Without a clear way to isolate the ionospheric seeing, we must settle 
on reasonable limits to its value.  The FWHM value of $23.2''$ derived
from fitted source sizes is a safe upper limit.  Being extra cautious, 
we take the lower limit to be 0.  We then use Equation \ref{size.eqn},
and apply the fitting errors on $\theta_{fit}$ \citep{1997PASP..109..166C} 
and the upper and lower limits to $\theta_{seeing}$ to calculate an upper 
and lower estimate to $\theta_{source}$.  We take the best estimate for 
$\theta_{source}$ to be the average of the upper and lower estimates, and
the error to be half the difference.  In this way, although we lack a 
definitive seeing measurement, the error bars on our source sizes 
encompass all reasonable values.  It is also worth noting that for sources
much larger than the restoring beam, the seeing adjustments are negligible.
Only for sources with sizes of order the beam size or smaller do the 
seeing adjustments make a significant difference.  Efforts are continuing
to devise an accurate way to measure the seeing effect in 74 MHz images.

\section{Completeness}
\label{com.sec}

Since our detection threshold is $5\sigma_{rms}$, and $95\%$
of sources will be measured to within $2\sigma_{rms}$, a reasonable 
estimate of the completeness level is $7\sigma_{rms}$.  Of course the
noise level varies over the field because we divided the map by 
the primary beam pattern to correct for its attenuation of sources.
Therefore, no single completeness level can apply to the entire maps. 
The map noise, $\sigma_{rms}$, rises from the field center to the 
edge, increasing from 0.8 mJy/beam to 1.6 mJy/beam at 325 MHz and 
from 55 mJy/beam to 110 mJy/beam at 74 MHz.  For the entire 
field, we can say we are complete at the 11.2 mJy and 770 mJy levels
at 325 and 74 MHz respectively, though these levels fall to 5.6 mJy
and 385 mJy for the inner regions of the fields.  That our measurements 
are consistent with the results of a previous survey 
\citep{1991PhDT.......241W} indicates that the theoretical values 
for completeness levels are reasonable.  

\section{Conclusions and Future Work}
\label{con.sec}

We have successfully surveyed the XMM-LSS region partially at 
325 MHz and completely at 74 MHz, detecting 256 and 211 objects
respectively.  We demonstrated that the full primary beam can 
be accurately imaged at sub-arcminute resolution at 74 MHz if 
ionospheric effects are numerically 
removed.  Realization of most of the scientific goals of this survey 
awaits comparison with upcoming optical and X-ray data from this field.  
However, the radio data alone have provided useful insight into 
the low-frequency radio source population, particularly at 74 MHz 
which has not previously been explored.  In the future we hope to
perform a more thorough survey of the XMM-LSS region, surveying a 
larger area at 325 MHz and to a deeper flux limit at 74 MHz. This 
will dramatically increase the number of radio sources we detect in 
this field, improving our ability to relate the nature and location
of extra-galactic radio sources to the environment they inhabit in 
the large scale structure of the universe.

\section{Acknowledgments}

The authors made use of the database CATS \citep{1997V} of 
the Special Astrophysical Observatory.  
This research has made use of the NASA/IPAC Extragalactic Database (NED)
which is operated by the Jet Propulsion Laboratory, Caltech, under contract
with the National Aeronautics and Space Administration.
Basic research in radio astronomy at 
the Naval Research Laboratory is supported by the office of Naval 
Research.

\newpage 

\begin{deluxetable}{ccccccccc}
\tabletypesize{\scriptsize}
\tablecaption{Statistical Overview of Ionospheric Refraction \label{table1}}

\tablehead{
\colhead{Source} &
\colhead{Source} &
\colhead{RA} &
\colhead{DEC} &
\colhead{Dist. to 3C63} &
\colhead{rms offset:} &
\colhead{First 200 min.} &
\colhead{rms offset:} &
\colhead{Remainder} \\
\colhead{\#} &
\colhead{Name} &
\colhead{($^\circ$)} &
\colhead{($^\circ$)} &
\colhead{($^\circ$)} &
\colhead{RA ($''$)} &
\colhead{DEC ($''$)} &
\colhead{RA ($''$)} &
\colhead{DEC ($''$)}
}
\startdata
1 & J0225.1$-$0035 & 36.284 & -0.592 & 1.7 & 9.7 & 13.3 & 4.0 & 4.5 \\
2 & J0228.8$-$0337 & 37.222 & -3.627 & 2.6 & 17.4 & 30.0 & 6.0 & 4.6 \\
3 & J0219.4$-$0539 & 34.868 & -5.663 & 3.7 & 23.2 & 44.9 & 4.7 & 8.8 \\
4 & J0235.1$-$0402 & 38.779 & -4.035 & 4.1 & 25.0 & 44.2 & 9.4 & 6.7 \\
5 & J0232.8$-$0645 & 38.225 & -6.765 & 5.7 & 32.7 & 56.1 & 9.6 & 11.1 \\
6 & not in table & 40.670 & -0.013 & 5.8 & 17.4 & 24.0 & 10.8 & 7.4 
\enddata
\end{deluxetable}

\newpage

\begin{deluxetable}{lrrrrrcrrrrr} 
\tabletypesize{\tiny}
\tablecaption{325 MHz Source List\label{table2}}  
\tablehead{
\colhead{Source} &
\colhead{$\alpha$ (J2000)} &
\colhead{$\sigma_{\alpha}$} &
\colhead{$\delta$ (J2000)} &
\colhead{$\sigma_{\delta}$} &
\colhead{Dist.} &
\colhead{T.} &
\colhead{$S_{int}$} &
\colhead{} &
\colhead{Size} &
\colhead{} &
\colhead{$\alpha$} 
\\
\colhead{} &
\colhead{(h m s)} &
\colhead{$('')$} &
\colhead{($^{\circ}$ $'$ $''$)} &
\colhead{$('')$} &
\colhead{$(^\circ)$} & &
\colhead{mJy} &
\colhead{major $('')$} &
\colhead{minor $('')$} &
\colhead{PA $(^\circ)$} &
\colhead{}
}
\startdata
J0219.5$-$0418 & 02 19 35.76 &    0.5 & $-$04 18 43.5 &    0.6 &   1.28 & S & $    29.0\pm    3.3$ & $\leq     5.1$ & $\leq    3.1$ & $   43\pm  12$ & $   -0.68$\\ 
J0219.7$-$0453 & 02 19 45.39 &    0.7 & $-$04 53 35.7 &    1.1 &   1.28 & S & $    25.8\pm    5.6$ & $   10.4\pm  2.5$ & $\leq    8.9$ & $    2\pm  11$ & $   -0.45$\\ 
J0219.9$-$0447 & 02 19 57.34 &    0.5 & $-$04 47 54.1 &    0.6 &   1.21 & S & $   137.6\pm    7.6$ & $    5.6\pm  0.2$ & $\leq    1.5$ & $    1\pm   1$ & $   -0.83$\\ 
J0220.2$-$0421 & 02 20 13.25 &    0.5 & $-$04 21 48.7 &    0.7 &   1.12 & S & $    20.1\pm    3.1$ & $    5.5\pm  1.3$ & $\leq    4.3$ & $  165\pm   10$ & $   -0.62$\\ 
J0220.4$-$0442 & 02 20 27.91 &    0.6 & $-$04 42 19.5 &    0.8 &   1.07 & S & $     9.9\pm    2.6$ & $\leq     8.8$ & $\leq    5.2$ & $  155\pm  18$ & $<   -0.94$\\ 
J0220.5$-$0450 & 02 20 31.91 &    0.6 & $-$04 50 02.6 &    0.9 &   1.08 & S & $    10.7\pm    2.8$ & $\leq     9.6$ & $\leq    5.1$ & $  176\pm  14$ & $<   -0.99$\\ 
J0220.5$-$0514 & 02 20 35.34 &    0.5 & $-$05 14 00.4 &    0.6 &   1.25 & S & $    50.0\pm    4.0$ & $    5.0\pm  0.6$ & $\leq    2.4$ & $  166\pm   4$ & $   -1.06$\\ 
J0220.6$-$0408 & 02 20 36.12 &    0.7 & $-$04 08 06.1 &    0.8 &   1.08 & S & $     9.6\pm    2.8$ & $\leq     7.9$ & $\leq    7.8$ & NA  & $<   -0.92$\\
\enddata
\end{deluxetable}

\newpage

\begin{deluxetable}{lrrrrrcrrrrr} 
\tabletypesize{\tiny}
\tablecaption{74 MHz Source List\label{table3}}  
\tablehead{
\colhead{Source} &
\colhead{$\alpha$ (J2000)} &
\colhead{$\sigma_{\alpha}$} &
\colhead{$\delta$ (J2000)} &
\colhead{$\sigma_{\delta}$} &
\colhead{Dist.} &
\colhead{T.} &
\colhead{$S_{int}$} &
\colhead{} &
\colhead{Size} &
\colhead{} &
\colhead{$\alpha$} 
\\
\colhead{} &
\colhead{(h m s)} &
\colhead{$('')$} &
\colhead{($^{\circ}$ $'$ $''$)} &
\colhead{$('')$} &
\colhead{$(^\circ)$} &
\colhead{} &
\colhead{Jy} &
\colhead{major $('')$} &
\colhead{minor $('')$} &
\colhead{PA $(^\circ)$} &
\colhead{}
}
\startdata
J0201.7$-$0419 & 02 01 45.21 &    5.5 & $-$04 19 24.8 &    3.6 &   5.71 & S & $    0.95\pm   0.29$ & $\leq    76.2$ & $\leq   26.2$ & $   62\pm   7$ & $   -0.66$\\ 
J0202.9$-$0605 & 02 02 56.33 &    3.2 & $-$06 05 22.5 &    2.2 &   5.64 & S & $    2.22\pm   0.33$ & $   44.5\pm 11.6$ & $\leq   24.1$ & $   66\pm   4$ & $   -0.66$\\ 
J0204.9$-$0655 & 02 04 54.31 &    3.7 & $-$06 55 10.5 &    3.5 &   5.48 & S & $    2.00\pm   0.37$ & $   50.6\pm 14.3$ & $\leq   37.0$ & $   46\pm   7$ & $   -0.44$\\ 
J0204.9$-$0715 & 02 04 54.90 &    6.8 & $-$07 15 56.4 &    3.1 &   5.64 & S & $    1.87\pm   0.42$ & $   67.8\pm 19.8$ & $\leq   38.4$ & $   78\pm   5$ & $   -0.67$\\ 
J0206.0$-$0511 & 02 06 03.21 &    2.3 & $-$05 11 30.5 &    2.3 &   4.69 & M & $    3.38\pm   0.41$ &    44.4 & &  & $   -0.74$\\ 
                            & 02 06 03.21 &    2.3 & $-$05 11 30.5 &    2.3 &   4.69 & C & $    1.61\pm   0.23$ & $\leq    47.3$ & $\leq   21.9$ & $   36\pm   6$ & $   -0.74$\\ 
                            & 02 06 06.11 &    4.0 & $-$05 11 40.3 &    3.2 &   4.68 & C & $    1.77\pm   0.32$ & $   54.8\pm 14.2$ & $\leq   33.7$ & $   55\pm   5$ & $   -0.74$\\ 
J0206.0$-$0503 & 02 06 04.29 &    3.0 & $-$05 03 35.9 &    2.2 &   4.67 & S & $    2.40\pm   0.33$ & $   45.4\pm 10.6$ & $\leq   37.3$ & $   73\pm   7$ & $   -0.59$\\
\enddata
\end{deluxetable}

\newpage

\begin{deluxetable}{lrrlll}
\tabletypesize{\scriptsize}
\tablecaption{Objects with known redshifts \label{table4}}

\tablehead{
\colhead{Object} &
\colhead{RA (J2000)} &
\colhead{DEC (J2000)} &
\colhead{Object Type} &
\colhead{Redshift} &
\colhead{Reference}
}
\startdata
J0210.8$-$0633 & 2 10 53.06 & $-$6 33 38.3   &     Galaxy   &     0.04340                & \citet{1999ApJ...523...87T} \\
J0213.7$-$0256 & 2 13 46.80 & $-$2 56 39.8   &     Galaxy   &     0.35680 $\pm$ 0.00020 & \citet{1995AJ....109...14O} \\
J0215.1$-$0702 & 2 15 07.65 & $-$7 02 06.4   &     QSO      &     2.22000                & \citet{2001ApJS..135..227B} \\
J0215.6$-$0222 & 2 15 41.98 & $-$2 22 52.0   &     QSO      &     1.17800                & \citet{1997MNRAS.284...85D} \\
J0217.4$-$0052 & 2 17 28.57 & $-$0 52 27.8   &     QSO      &     2.45476 $\pm$ 0.00165 & \citet{2001SDSSe.1...0000:} \\
J0219.1+0120 & 2 19 07.25 & +1 20 58.3   &     QSO      &     1.62300                & \citet{1992NED11.R......1N} \\
J0220.5+0027 & 2 20 33.43 & +0 27 57.7   &     Galaxy   &     0.27226 $\pm$ 0.00016 & \citet{2001SDSSe.1...0000:} \\
J0220.9$-$0156 & 2 20 54.26 & $-$1 56 52.5   &     Galaxy   &     0.17500                & \citet{1991ApJS...75..297H} \\
\enddata
\end{deluxetable}

\end{document}